\documentclass[a4paper,fleqn]{cas-dc}

\usepackage[authoryear,longnamesfirst]{natbib}

\usepackage{color,xcolor}

\usepackage{blkarray}
\usepackage{booktabs}
\usepackage{gensymb}
\usepackage{algorithm}
\usepackage{algpseudocode}

\usepackage{caption}
\usepackage{subcaption}

\def\tsc#1{\csdef{#1}{\textsc{\lowercase{#1}}\xspace}}
\tsc{WGM}
\tsc{QE}

\newcommand{\blue}[1]{\textcolor{blue}{#1}}

\begin{document}
\let\WriteBookmarks\relax
\def\floatpagepagefraction{1}
\def\textpagefraction{.001}

\shorttitle{Gyroscope Calibration Method Without External Device}

\title [mode = title]{In-Field Gyroscope Autocalibration with Iterative Attitude Estimation}

\author[inst1, inst2, inst3]{Li Wang}

\affiliation[inst1]{organization={Shandong Provincial Hospital Affiliated to Shandong First Medical University \& Shandong Academy of Medical Sciences},
            city={Jinan},
            postcode={250000}, 
            state={Shandong},
            country={China}}
            
\affiliation[inst2]{organization={College of Artifical Intelligence and Big Data for Medical Sciences, Shandong First Medical University \& Shandong Academy of Medical Sciences},
            city={Jinan},
            postcode={250000}, 
            state={Shandong},
            country={China}}
            
\affiliation[inst3]{organization={CSIRO Space and Astronomy},
            addressline={P.O. Box 1130}, 
            city={Bentley},
            postcode={6102}, 
            state={WA},
            country={Australia}
            }
            
\affiliation[inst4]{organization={Faculty of Engineering and IT, University of Technology Sydney},
            addressline={81 Broadway}, 
            city={Ultimo},
            postcode={2007}, 
            state={NSW},
            country={Australia}
            }

            
\affiliation[inst5]{organization={Faculty of Health, University of Technology Sydney},
            addressline={235 Jones St}, 
            city={Ultimo},
            postcode={2007}, 
            state={NSW},
            country={Australia}
            }
            
\affiliation[inst6]{organization={School of Computer, Data and Mathematical Sciences, Western Sydney University},            addressline={Second Ave}, 
            city={Kingswood},
            postcode={2747}, 
            country={Australia}}

\author[inst5]{Rob Duffield}           
\author[inst5]{Deborah Fox}
\author[inst5]{Athena Hammond}
\author[inst4]{Andrew J. Zhang}
\author[inst6]{Wei Xing Zheng}
\author[inst2, inst4]{Steven W. Su}
\cormark[1]

\begin{abstract}
This paper presents an efficient in-field calibration method tailored for low-cost triaxial MEMS gyroscopes often used in healthcare applications. Traditional calibration techniques are challenging to implement in clinical settings due to the unavailability of high-precision equipment. Unlike the auto-calibration approaches used for triaxial MEMS accelerometers, which rely on local gravity, gyroscopes lack a reliable reference since the Earth's self-rotation speed is insufficient for accurate calibration. To address this limitation, we propose a novel method that uses manual rotation of the MEMS gyroscope to a specific angle (360°) as the calibration reference. This approach iteratively estimates the sensor's attitude without requiring any external equipment. Numerical simulations and empirical tests validate that the calibration error is low and that parameter estimation is unbiased. The method can be implemented in real-time on a low-energy microcontroller and completed in under 30 seconds. Comparative results demonstrate that the proposed technique outperforms existing state-of-the-art methods, achieving scale factor and bias errors of less than $2.5\times10^{-2}$ for LSM9DS1 and less than $1\times10^{-2}$ for ICM20948.

\end{abstract}




\maketitle

\section{Introduction}
\begin{figure*}
     \centering
     \begin{subfigure}{0.177\textwidth}
         \centering
         \includegraphics[width=\textwidth]{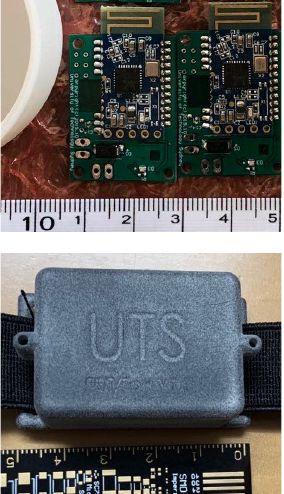}
         \caption{Designed wearable motion tracking device.}
         \label{IMUFIG}
     \end{subfigure}
     \begin{subfigure}{0.6\textwidth}
         \centering
         \includegraphics[width=\textwidth]{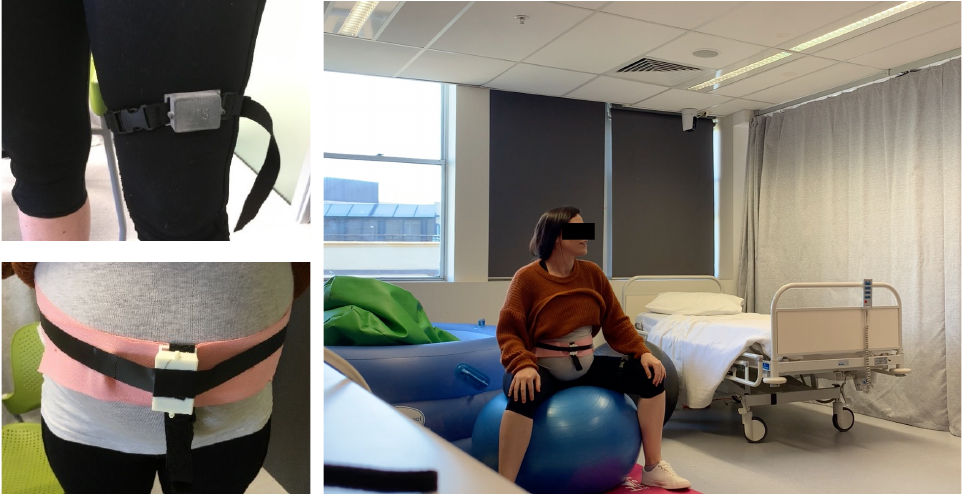}
         \caption{The mobilisation monitoring of woman experiencing high risk pregnancy in a pilot study. The devices are attached on the left thigh and in front of the abdomen.}
         \label{RealExp}
     \end{subfigure}
     \hfill
        \caption{The designed motion tracking device and its application in a pilot study. }
        \label{fig:usage}
\end{figure*}

The gyroscope is essential equipment for measuring angular velocity in a wide range of technologies, such as motion tracking \citep{SCAPELLATO2005418,MotionTracking}, vibration measurement \citep{yu2017single,halkon2021establishing}, and wearable health monitoring \citep{healthmonitor,petropoulos2020wearable,farooq2022comprehensive}. Recently, we designed a low-cost micro-electromechanical (MEMS) inertial measurement unit (IMU) device to be worn by preganant women during childbirth to explore the impact of hospital birth room configuration upon mobility for women with complex pregnancies (See Figure. \ref{fig:usage}). However, a key concern of measurement in such ecological settings is that the accuracy of low-cost gyroscopes are usually low. For instance, when calculating the attitude, the integration will lead to the accumulation of drift and scale factor error \citep{accumulate}. In addition, owing to the poor repeatability and significant volatility, on every booting or under different environmental conditions, the scale factor and biases change \citep{errorTemperature, zhang2017thermal}. Therefore, it is necessary to calibrate the gyroscope on each initialization or when environmental conditions change. However, calibration processes are normally time-consuming or cost ineffective, and thus in field or clinical context, practitioners need frequent calibration of the gyroscope, which needs to be a simple and efficient process.

\subsection{Preliminary Study}\label{sec:preliminary}
The development of the in-field wearable sensor calibration method is motivated by the transdisciplinary project 'The use of wearable technologies to explore the impact of hospital birth room configuration upon mobility in childbirth for women with complex pregnancies' under the support of the Faculty of Engineering and Information technology Cross-Faculty Collaborative Scheme, at the University of Technology Sydney (UTS).

The study aimed to monitor the movements of high-risk pregnant women during labor.  In order to protect the privacy of potential participating birthing women in future studies, the device could not use the conventional video-camera based tracking system. Instead, motion-tracking technology based on the integration of inertial sensors and wireless sensing was developed and pre-tested. One of the subsystems is a wearable module that includes wireless IMU for gait monitoring and assisted behaviour recognition.

\begin{figure}
	\centering
		\includegraphics[height=0.35\textwidth]{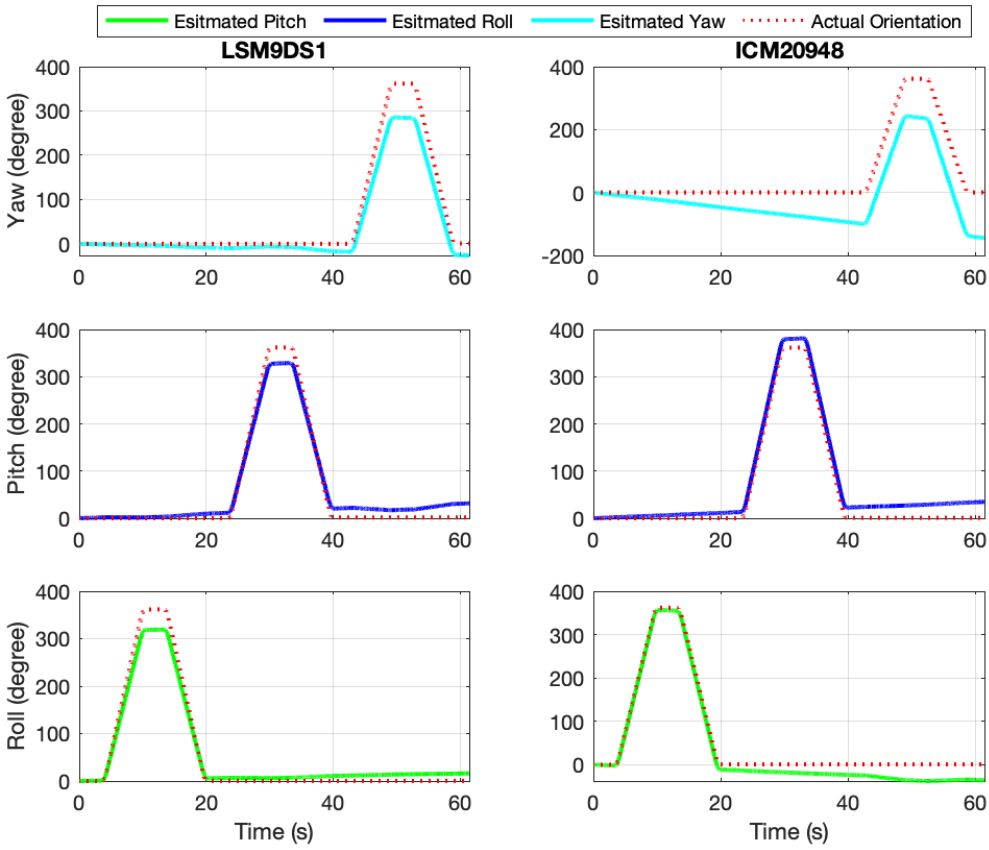}
	\caption{Estimated orientation using raw gyroscope readings before calibration from two models of the gyroscope. Left: LSM9DS1. Right: ICM20948.}
	\label{oriBefore}
\end{figure}

To test the accuracy of the developed portable IMU device, especially the gyroscope, we implemented the orientation angle estimation tests. The gyroscope to be tested is placed on a 3-axis turntable, which rotates 360 degrees clockwise along the x-axis and then 360 degrees counterclockwise. We repeat this action on the y and z axes. Figure. \ref{oriBefore} indicates that the angle estimation error is significant, which will potentially influence the gait estimation and gesture analyses.

To improve the estimation accuracy, we tried to develop in-field calibration methods for both accelerometers and gyroscopes. Despite the extensive amount of literature investigating in-field calibration of triaxial accelerometers or magnetometers \citep{triaxialMathematicalModel,ye2017efficient,gao2017accelerometers,sipos2011analyses,wu2021calibration}; few studies discuss the calibration of the gyroscope under non-laboratory environments. This gap exists because local gravity serve as direct calibration references for accelerometers. However, when it comes to gyroscopes, the Earth's self-rotation speed is insufficient as a standalone reference for accurate calibration. In this study, we introduce a novel, easy-to-use method for gyroscope calibration that significantly improves angle estimation accuracy, as demonstrated in subsequent discussions.

\subsection{Existing Gyroscope Calibration Methods}

Existing research recognized the critical role played by the methods and accuracy of gyroscope calibration. In particular, the conventional calibration method uses a turntable to provide a standard rotation speed for the gyroscope \citep{truntable}. This method can provide high calibration accuracy but requires expensive and precise instruments and complex calibration procedures that preclude its use in consumer electronic devices or clinical settings. Several recent studies \citep{camera,mag,acc1,acc2,lu2022all,xia2022estimation} proposed several calibration methods that do not need precision equipment. Specifically, in \citep{camera}, a camera-aided calibration method was reported. The images provide the orientation and position information of the sensor to confirm its orientation, resulting in a high computational complexity. In a separate method \citep{mag}, a homogeneous magnetic field was employed as the calibration reference. The natural geomagnetic field is very weak and is easily affected by the alternating electric field, again making it difficult to implement outside the laboratory. In \citep{acc1,acc2}, an accelerometer-aided gyroscope calibration method was presented. The accelerometer was first calibrated using the multi-position method. Then, the rotation speed of the sensor body is provided by the accelerometer. In this method, an extra triaxial accelerometer with the same coordinate system is needed, and the calibration error of the accelerometer affects the calibration accuracy of the gyroscope. Besides, the calibration process takes more than ten minutes. 
In \citep{lu2022all}, a two-step method was proposed to estimate all parameters for a six-axis IMU. A turntable was employed to perform a 12-points rotation, and a Kalman filter was designed to compensate for the parameter error.
However, from the above discussion, we can infer that these gyroscope calibration methods relying on external equipment are unsuitable for scenarios where the external calibration device is unavailable, such as in a field or clinical setting akin to a busy and chaotic hospital birthing room. Besides, an accelerometer may not necessarily be integrated into the device and cannot provide calibration reference for a gyroscope.
This paper, for the first time, introduced an easy-to-get rotation angle that can be used as the calibration reference. In this case, the gyroscope calibration no longer requires a constant rotation speed or reference from an accelerometer, thereby getting rid of the shackles of high-precision external equipment.

\subsection{Summary of Our Contributions}
The contributions of this paper can be summarized as follows: 
\begin{itemize}
    \item We proposed an autocalibration method for triaxial gyroscopes. This calibration method is implemented in a microcontroller and only takes 30 seconds without using any external device.
    \item We proposed a practical but accurate method to provide the calibration reference and validated the effectiveness of the proposed calibration method in both numerical simulation and real-time experiments. 
    \item We analysed the sources of errors and proposed the potential solutions for eliminating the error. 
    \item We designed a low-energy, cost-effective, and wearable wireless movement-tracking device for health monitoring. We applied the proposed calibration approach in the wearable device and significantly improved angle estimation accuracy.
\end{itemize}

\section{Methodology}\label{sec:method}

\begin{figure}[h]
    \centering
	\includegraphics[height=0.3\textwidth]{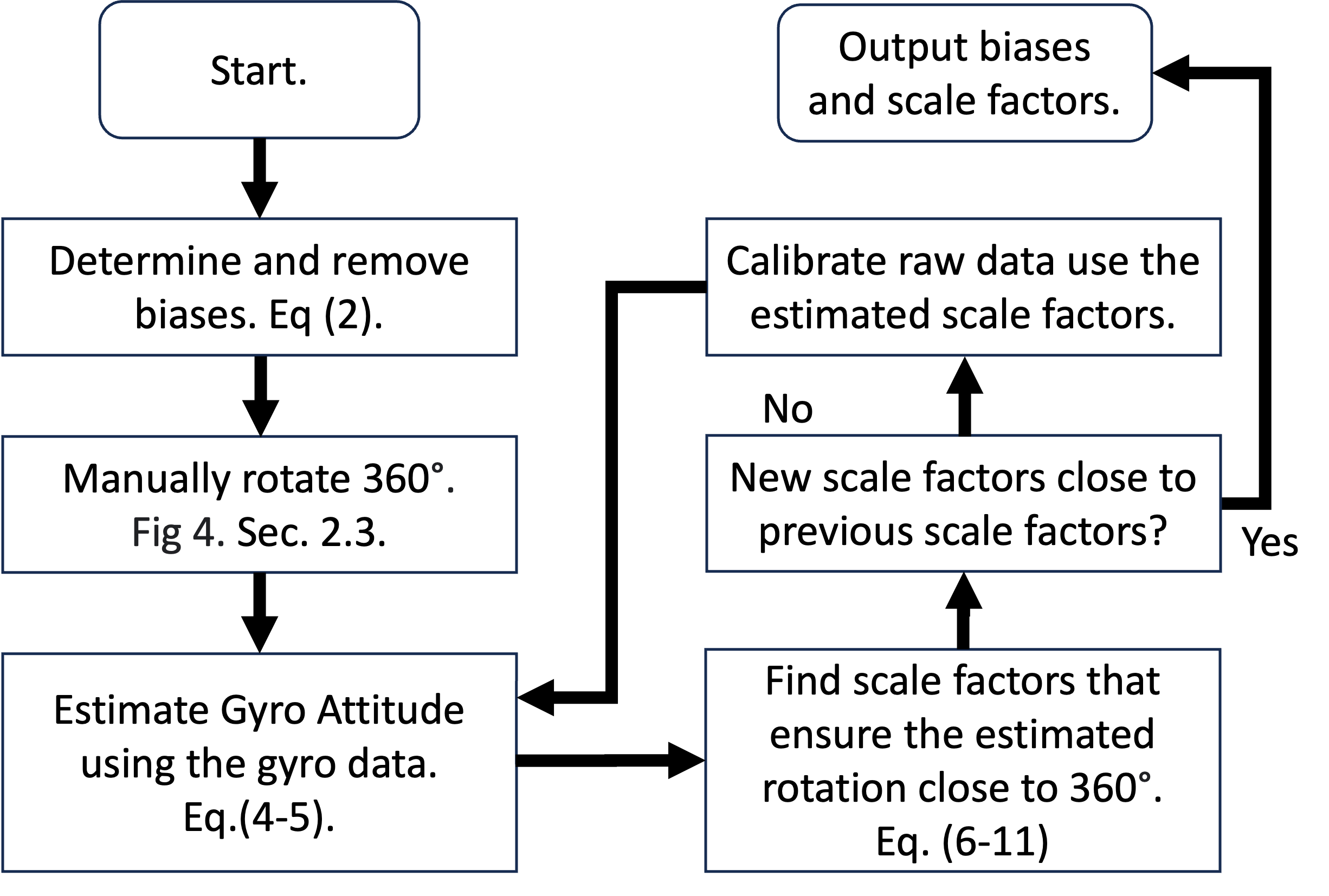}
	\caption{Overview of the proposed method.}
	\label{fig:overview}
\end{figure}

In this section, we first propose an autocalibration method that uses manual rotation angle as a calibration reference. Then, a four-observation calibration method is reported. \blue{The overview of the proposed celebration method is shown in Figure. \ref{fig:overview}.} In this study, we ignored the earth's rotation as it is often submerged in the noise of the low-cost gyroscope.

\subsection{Notations and data formulations}
For a triaxial gyroscope, at time $t$, the measurements contains three elements $M_t=[m_{x,t},m_{y,t},m_{z,t}]^T$. For a period of measurement of length $T$, we have $M_t, t \in [0,T]$. Divide $T$ into $m$ stages and get $S_0 = [0,s_0]$, $S_1 = [s_0,s_1], \dots ,S_m = [s_{m-1},s_m]$.

\subsection{Calibration principle}
Numerous parameters contribute to errors in gyroscopes, including the scale factor, biases, orthogonality errors, and angular rate sensitivity. However, in healthcare applications, both acceleration and rotation speeds are generally low. As a result, orthogonality errors and acceleration sensitivity often become unobservable and can be neglected. Therefore, this study focuses on a simplified 6-parameter calibration model to estimate the unknown parameters. Therefore, the relationship between the actual angular velocity components $G_t^b=[g^b_{x,t}, g^b_{y,t}, g^b_{z,t}]^T$ and the measured angular velocity $M_t=[m_{x,t},m_{y,t},m_{z,t}]^T$ at $i$ th rotation are described as:
\begin{equation}\label{model}\begin{bmatrix}
g^b_{x,t} \\g^b_{y,t} \\g^b_{z,t}
\end{bmatrix}=\begin{bmatrix}
k_x & 0 & 0 \\0 & k_y & 0 \\0 & 0 & k_z
\end{bmatrix} \begin{pmatrix}\begin{bmatrix}
m_{x,t} \\m_{y,t} \\m_{z,t}
\end{bmatrix} + \begin{bmatrix}
b_x \\b_y \\b_z
\end{bmatrix}\end{pmatrix},\end{equation}
where the $k_x,k_y,k_z$ and $b_x,b_y,b_z$ represents the scale factors and biases, respectfully. \blue{In this paper, nonlinearity is disregarded because modern MEMS IMUs typically have nonlinearity on the order of thousandths \citep{li2024site, ICM20948Datasheet}, which is negligible compared to the scale factor and bias errors considered. }

When the gyroscope is in a static state ($S_0$), we can remove the biases by subtracting the average gyroscope readings of each axis as follows:
\begin{equation}\label{eq:biasest}
    \left\{
        \begin{array}{l}
            b_x=-\frac{1}{s_0}\sum_{t=0}^{s_0} m_{x,t}\\
            b_y=-\frac{1}{s_0}\sum_{t=0}^{s_0} m_{y,t}\\
            b_z=-\frac{1}{s_0}\sum_{t=0}^{s_0} m_{z,t}\\
        \end{array}
    \right., t\in S_0
\end{equation}
where $s_0$ is the end time of the stationary stage. After the biases are compensated, Eq. \eqref{model} becomes:

\begin{equation}\label{modelsimple}
\begin{bmatrix}
g_{x,t}\\ 
g_{y,t}\\ 
g_{z,t}
\end{bmatrix}= \begin{bmatrix}
k_x m_{x,t}\\ 
k_y m_{y,t}\\ 
k_z m_{z,t}
\end{bmatrix}
\end{equation}

When the sampling time $\Delta t$ is small, the rotation angle accrued during a sampling period is also typically small. Under these conditions, the rotational angle can be approximately treated as a vector. We initially estimate the orientation of the gyroscope using the gyroscope readings $M_t=[m_{x,t},m_{y,t},m_{z,t}]^T$.
The angle of rotation is $\theta_t=\|\Delta t M_t\|$ and the unit vector for the axis of rotation is $U_t=\frac{\Delta t M_t}{\theta_t}$.
Based on Rodrigues' formula, we can transfer the measurement to rotation matrix. The rotation matrix can be written concisely as
\blue{
\begin{equation}\label{eq:rotationmat}
    R_t=(\cos\theta_t)I+(\sin\theta_t)[U_t]_x+(1-\cos\theta_t)U_t U_t^T,
\end{equation}}
where $I$ is a 3-by-3 identity matrix, and $[U_t]_x$ is the cross product matrix. We consider the orientation of the gyroscope to start reading (i.e. $t=0$) as the initial frame ($^n$), and the gyroscope as body frame ($^b$). Then, the rotation matrix $C^n_{b,t}$ which transforms the body frame to initial frame can be calculated as follows:
\begin{equation}
    C^n_{b,t}=
    \prod_{0}^{t}R_t
\end{equation}

Then, we transforms the rotation vectors in body frame to initial frame. The rotation vectors in the initial frame are:
\begin{equation}\label{rotvecini}
    G^n_t=C^n_{b,t}G^b_t
\end{equation}
Substitute Eq. \eqref{modelsimple} into Eq.\eqref{rotvecini}, we have,
\begin{equation}\label{rotvecinivar1}
    G^n_t=C^n_{b,t}(KM_t),
\end{equation}
\blue{where $K=\mathrm{diag}[k_x,k_y,k_z]$ contains the scale factors.} Let 
\begin{equation}
\label{eq:tildeC}
    \tilde{C^n_t}=\begin{bmatrix}
c_{11,t}m_{x,t} & c_{12,t}m_{y,t} & c_{13,t}m_{z,t}\\ 
c_{21,t}m_{x,t} & c_{22,t}m_{y,t} & c_{23,t}m_{z,t} \\ 
c_{31,t}m_{x,t} & c_{32,t}m_{y,t} & c_{33,t}m_{z,t} 
\end{bmatrix},
\end{equation}
then, Eq. \eqref{rotvecinivar1} becomes,
\begin{equation}
    G^n_t=\tilde{C^n_t} K,
\end{equation}
Note that $\tilde{C^n_t}$ is a constant matrix as all entries of $\tilde{C^n_t}$ are from in $C^n_{b,t}$ and $M_t$.

\begin{figure}
	\centering
		\includegraphics[width=0.45\textwidth]{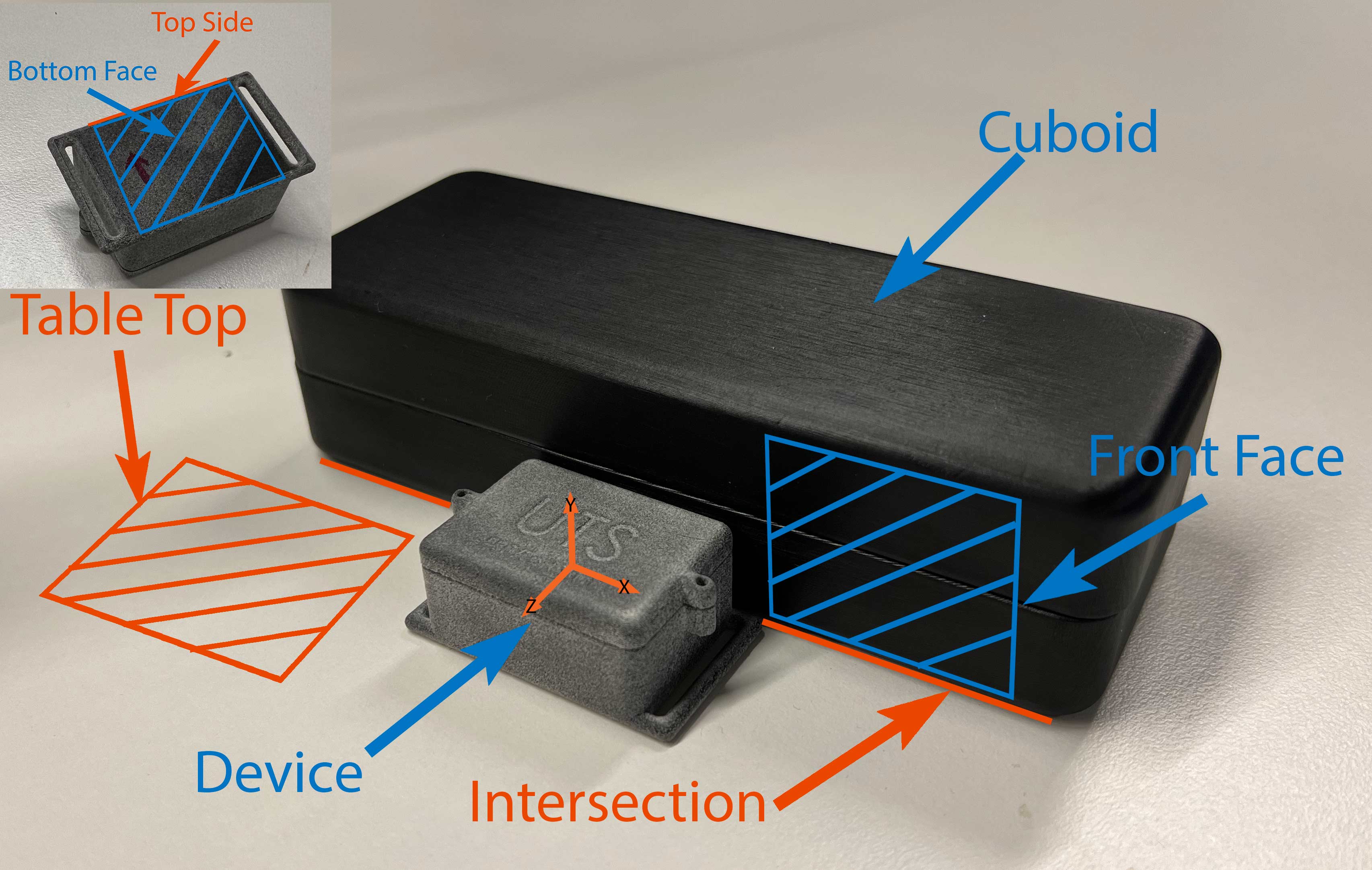}
	\caption{Experimental system for gyroscopes calibration. Main figure: the initial position of the device. Top left figure: bottom view of the device.}
	\label{system}
\end{figure}

Next, we calculate average rotation speed using the rotation vectors in initial frame. 
\begin{equation}\label{eq:10}
    \Bar{G}^n=\frac{1}{s_m-s_{m-1}}\sum_{t=s_{m-1}}^{s_m} G_t^n=\frac{1}{s_m-s_{m-1}}\sum_{t=s_{m-1}}^{s_m} \tilde{C}_t^n K
\end{equation}
The main idea of the proposed method is that we use two different methods to calculate the average rotation speed, i.e., the method based on \eqref{eq:10} and the method based on the total rotation angle divided by rotation time. 
The modulus of average rotation speed calculated by \eqref{eq:10} should equal to the actual modulus of average rotation speed,
\begin{equation}\label{eq:final}
\begin{split}
    \Bar{G}^{nT} \Bar{G}^n =& (\frac{1}{s_m-s_{m-1}}\sum_{t=s_{m-1}}^{s_m} \tilde{C}_t^n K)^T\\ &\times  (\frac{1}{s_m-s_{m-1}}\sum_{t=s_{m-1}}^{s_m} \tilde{C}_t^n K)
    =  (\frac{\theta_{total}}{t})^2 ,
\end{split}
\end{equation}
where $\theta_{total}$ is a known value, $\tilde{C}_t^n$ is a constant value calculated by \eqref{eq:tildeC}, and $t$ is the time of collecting $N$ samples, and $\theta_{total}$ is the total rotation angle performed in each experiment. 
For an embedded system, time $t$ can be measured precisely. Turning a precise angle in the experiment becomes the key to the success of the experiment.

\subsection{Getting the Precise Rotation Angle.}\label{sec:process}

\begin{figure}
	\centering
		\includegraphics[width=0.35\textwidth]{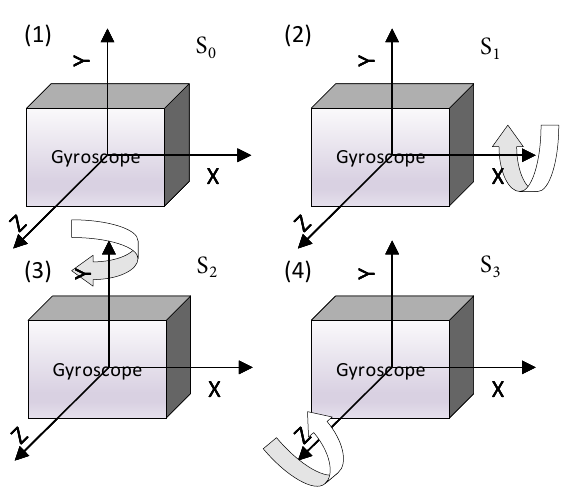}
	\caption{4-observations rotation protocol for gyroscope calibration. (1) Stationary stage. (2)-(4) Rotating stage: Manually rotate the gyroscope 360 degrees clockwise along the x,y,z axes.}
	\label{rotationmethod}
\end{figure}

In this study, a very simple principle is used to provide an accurate rotation angle, which is that the total rotation angle of the object after rotated one cycle is 360 degrees.
Specifically, we use the table top and the front face of a cuboid placed on the table to define initial position of the device (see Figure. \ref{system}). It should be noted that when the bottom face of the device coincides with the table top, and the top side of the device bottom face coincides with the intersection of the desktop and the front face of the cuboid, the posture of the device relative to the desktop is uniquely determined. The two planes are not required to be strictly vertical, but they are required to be stable prior to performing rotations outlined in Figure. \ref{rotationmethod}. 
First, the device is in its initial position on the table top before gently rotating 360 degrees around the x-axis of the gyroscope. Then, we put the gyroscope back to its initial position, which is deemed one rotation. This process is repeated in the y and z-axes to obtain a total of 3 rotational observations. The gyroscope stays still for three seconds between each rotation. It is worth noting that high-accuracy equipment and customized case are not used in this process.
It should be emphasized that maintaining uniform rotation speed and direction during the experiment is crucial for reducing calibration errors attributable to non-linearities in the scaling factors. However, we will later demonstrate through both simulations and empirical tests that the calibration accuracy remains acceptable even in the presence of modest variations in rotation speed and direction, e.g. in a manual rotation scenario.

Overall, the calibration process includes four stages, including one stationary stage and three rotation stages. The different stages can be identified by finding the pause between each rotation.

\subsection{Iterative calibration method.}
\begin{algorithm}
\caption{Iterative calibration algorithm}\label{alg:dist}
\begin{algorithmic}
\Require Gyroscope readings $M_t=[m_{x,t},m_{y,t},m_{z,t}], i \in [0,T]$ for the stationary stage ($S_0$) and three rotating stages ($S_1,S_2,S_3$) (See Figure. \ref{rotationmethod}). $i=1$, $K^0=[1,1,1].$ 
\Ensure $k_x,k_y,k_z,b_x,b_y,b_z$.
\State 1. Calculate biases ($b_x,b_y,b_z$) with Eq. \ref{eq:biasest} during. The compensated measurement is $\hat{M_t}, t \in [0,T]$.
\State 2. Calculate scale factors, $K^i$, using Eq. \eqref{modelsimple}-\eqref{eq:final}. 
\State 3. Compensate $M_t$ with $K^i$, i++.
\State 4. \textbf{Repeat} 2. and 3. \textbf{until} $\|(K^{i-1}-K^{i-2})\| \leq 10^{-5}$
\end{algorithmic}
\end{algorithm}

When calculating the rotation matrix in \eqref{eq:rotationmat}, the raw gyroscope measurement is used. This makes the transformation matrix $C^n_{b,t}$ also contain measurement errors. To eliminate this error, we use the result of the first calibration result as the initial value to compensate the gyroscope reading. The calibration process iterates until the scale factor no longer changes. The overall process of calibration is summarized in Algorithm \ref{alg:dist}.

\section{Simulation}\label{sec:simulation}

To validate the proposed method under different conditions such as scale factors, biases, mounting misalignment, and measurement noise, we first inspect the proposed method using simulations. For the purpose of simulating under the similar hardware condition, the following assumptions on the parameter are given, and the results are explained thereafter:
\begin{enumerate}
    \item The scale factors and biases follow uniform distributions $U(80\%,120\%)$ and $U(-5\degree/s,5\degree/s)$, respectively. The parameters of low-cost MEMS gyroscopes are usually in between these values.
    \item The mounting misalignment on each axis follows a uniform distribution $U(-10\%,10\%)$. In the absence of high-precision fixtures, it is almost impossible to mount the gyroscope accurately.
    \item The measurement noise is assumed following a Gaussian distribution with zero mean and two different noise $ \mathcal{N}(0,\,0.03^2)$ and $ \mathcal{N}(0,\,0.15^2)$. The typical rate noise density is between $0.003-0.015\degree/sec/\sqrt{Hz}$. At a $100Hz$ sampling frequency, we consider the standard deviation of root-mean-square noise as $0.03-0.15 \degree/sec$.
\end{enumerate}

\begin{figure}
	\centering
		\includegraphics[height=0.35\textwidth]{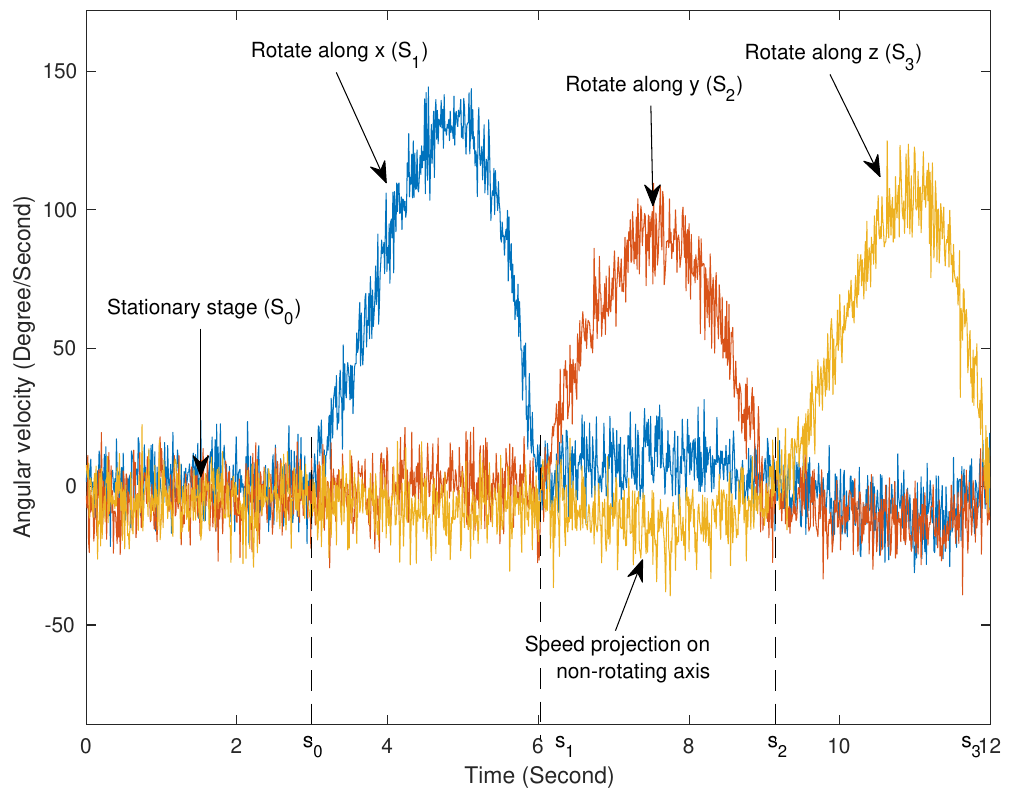}
	\caption{Typical simulated measurements of the proposed calibration method under $0.15 \degree/sec$ noise level. Speed variation and speed projection on non-rotating axis were to simulate the manual rotation process.}
	\label{simulationrun}
\end{figure}

Based on the above assumptions, we formulate the following simulation process. 30 sets of parameters are randomly generated based on the assumptions above. For each set of parameters, 500 simulations were repeated. During each simulation, we created a 4-observation measurement according to the experiment scheme described in Section \ref{sec:process}. For the purpose of showing that the proposed method is not sensitive to the speed variation, the rotation speed in the rotating stage was represented by a randomly generated Bézier curve \citep{Beziercurve}. A typical run of one simulation is shown in Figure. \ref{simulationrun}. We also generated a testing set for each simulation. The gyroscope were assumed to be rotating constantly and have equal speed projection on each axis during testing process.

After obtaining the 4-observation measurement,  Algorithm \ref{alg:dist} is employed to calculate the biases and scale factors. We used the estimated biases and scale factors to compensate the test set and compare with the actual value. 15000 simulations were conducted for testing our efficient calibration method. 

\begin{figure}
	\centering
		\includegraphics[height=0.35\textwidth]{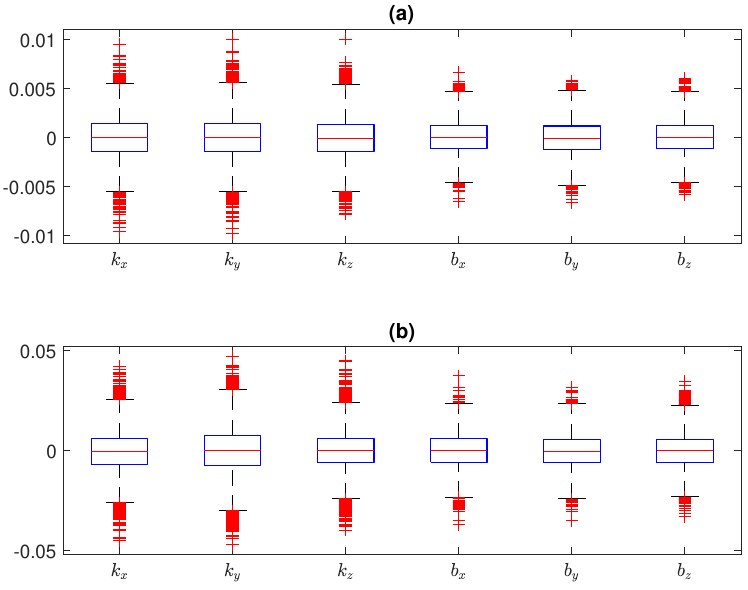}
	\caption{The error of estimated biases and scale factors compared with the true values at different noise levels. (a)$0.03 \degree/sec$ noise level. (b)$0.15 \degree/sec$ noise level.}
	\label{errorboxplot}
\end{figure}

We first observe the error between the estimated parameters and the actual parameters. Box plot Figure. \ref{errorboxplot} was used to analysis the statistical characteristics of the error. The results show that the estimated scale factors and bias terms are unbiased with a zero median value of the estimation error. The majority of the estimation error are within $\pm 5.5\times10^{-3}$ for $0.03 \degree/sec$ noise level and within $\pm 2.5\times10^{-2}$ for $0.15 \degree/sec$ noise level. Besides, each subplot in Figure. \ref{errorgyroboxplot} shows statistical characteristics of the mean error of the calibrated measurement value and the actual value in the testing set in 15,000 simulations. The error is defined as:
\begin{equation}
    e_l=\frac{\sum^N_{i=1}(g^t_{l,t}-\hat{m}^t_{l,t})}{N}, l=x,y,z,
\end{equation}
where $g^t$ and $\hat{m}^t$ stands for the true value and calibrated measurement value in the test set. Under both noise levels, the error after calibration is significantly reduced compared to the error before calibration. Among them, under the noise level of $0.03 \degree/sec$, the measurement error is reduced to 3\% of the previous.

\section{Experiments}\label{sec:experiment}

In this section, we empirically implemented and validated the proposed method. For the purpose of showing the efficiency and accuracy of our method, we compared the proposed method with three existing methods. 
\subsection{Experiments Device and Hardware Design}
The developed wearable MEMS IMU device is shown in Figure. \ref{IMUFIG}.  This device is composed of a 3D printed case, a 9-degree-of-freedom inertial sensor (ICM20948), an RF transceiver (RTL8762AG) and a microcontroller (STM32F103T8U6). Owing to the low energy consumption features of the chosen components and adjustable inertial sensor sampling frequency ($4.4Hz-562.5Hz$), the device can continuously work for 14 hours with a 400 mAh lithium polymer battery. Besides, to further reduce power consumption, we use Bluetooth Low Energy technology to transmit the collected data to the host computer. In addition, the 3D-printed case has good airtightness, which makes the device meet the IP67 waterproof standard. For ease of use, the device is also equipped with a wireless charging function. 

\begin{figure}
	\centering
		\includegraphics[height=0.35\textwidth]{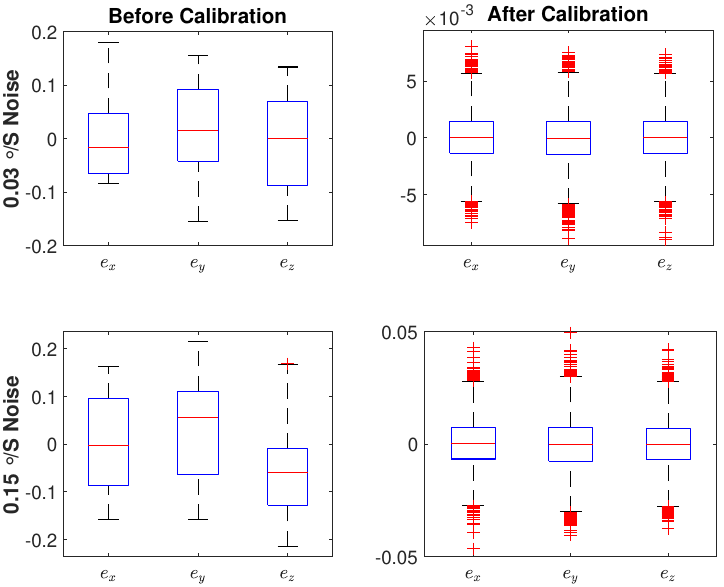}
	\caption{The mean error of the calibrated angular velocity compared with the actual value of each axis. Row: under different noise levels. Column: before and after calibration.}
	\label{errorgyroboxplot}
\end{figure}

\subsection{Experimental Setting}
Two commercial-grade low-cost gyroscopes are tested in this study. One is LSM9DS1 from ST-Microcontroller, and another one is ICM20948 from TDK InvenSense. 
The LSM9DS1 was set to $104 Hz$ sampling rate with a full-scale range of $245 \degree/sec$. On ICM20948, the sampling frequency was set to $104Hz$, and the full-scale range was set to $250 \degree/sec$. The parameters are summarized in Table. \ref{relatepara}. The room temperature was set to 22\degree C. All devices were preheated for 3 minutes before experiments.
\begin{table}[h]
  \centering 
  \caption{Related parameters of test gyroscope}\label{relatepara}
    \begin{tabular}{ccc}
      \toprule
      Parameter & LSM9DS1   & ICM20948 \\
      \hline
      Sensitivity Scale Factor Tolerance & Not Provide & $\pm4.5\%$\\
      Zero-Rate Output Tolerance (dps)& $\pm30$ &	$\pm6.25$\\
      Scale range (dps) & 245&250\\
      Sampling Frequency (Hz) &104 &104\\
      \hline \\
    \end{tabular}
\end{table}

\begin{table*}[t]
  \centering
  \caption{Calibration results comparison of LSM9DS1.}\label{resultsL9}
  \resizebox{0.8\textwidth}{!}{%
    \begin{tabular}{ccccc}
      \toprule
      Parameter & Results of proposed   & Results of conventional  &Results of accelerometer& Results of servomotor \\
      & calibration method &turntable method \citep{truntable}& -aided method \citep{acc2}&-aided method \citep{servomotor} \\
      \hline
      $k_x$             &   1.2374  (0.0245)    &   1.2619  &   1.2464 (0.0154)         &1.2590 (0.0029)\\
      $k_y$             &   1.1764  (0.0202)    &	1.1966  &   1.1806 (0.0160)         &1.2030 (0.0064)\\
      $k_z$             &   1.1504  (0.0032)    &   1.1536  &   1.1603 (0.0067)         &1.1441 (0.0095)\\
      $b_x(\degree/s)$  &   -3.1065 (0.0215)    &   -3.1275 &   -3.1156 (0.0124)        &-3.1436 (0.0157)\\
      $b_y(\degree/s)$  &   1.6171  (0.0141)    &	1.6312  &   1.6137  (0.0175)        &1.6585 (0.0274)\\
      $b_z(\degree/s)$  &   -1.4467 (0.0188)    &   -1.4655 &   -1.4794 (0.0139)        &-1.4493 (0.0162)\\
      Time (Second)     &   27                  &   1533    &   570                     &   60\\
      Equipment         &   N/A                 &High-precision turntable &Accelerometer&Servomotor\\
      \hline 
      \multicolumn{5}{l}{Notes. The value in parentheses is the absolute error compared with the turntable method.}\\
    \end{tabular}%
    }
\end{table*}

\begin{table*}[]
  \centering
  \caption{Calibration results comparison of ICM20948}\label{resultsICM}
  \resizebox{0.8\textwidth}{!}{%
    \begin{tabular}{ccccc}
      \toprule
      Parameter & Results of proposed   & Results of conventional  &Results of accelerometer& Results of servomotor \\
      & calibration method &turntable method \citep{truntable}& -aided method \citep{acc2}&-aided method \citep{servomotor} \\
      \hline
      $k_x$             & 0.9859 (0.0037)   & 0.9822    & 0.9745 (0.0077) &0.9780 (0.0042)\\
      $k_y$             & 1.0096 (0.0088)   & 1.0183    & 1.0189 (0.0006) &1.0168 (0.0015)\\
      $k_z$             & 0.9710 (0.0078)   & 0.9788    & 0.9879 (0.0091) &0.9765 (0.0023)\\
      $b_x(\degree/s)$  & -0.5236 (0.0067)  & -0.5169   & -0.5159 (0.0011) &-0.5250 (0.0081)\\
      $b_y(\degree/s)$  & -1.4347 (0.0055)  & -1.4402   & -1.4488 (0.0086) &-1.4473 (0.0071)\\
      $b_z(\degree/s)$  & 1.0525 (0.0023)   & 1.0502    & 1.0537  (0.0035) &1.0566 (0.0064)\\
      Time (Second)     &   29              &   1495    &   565             &   60\\
      Equipment         &   N/A                 &High-precision turntable &Accelerometer&Servomotor\\
      \hline
      \multicolumn{5}{l}{Notes. The value in parentheses is the absolute error compared with the turntable method.}\\
    \end{tabular}%
    }
\end{table*}

\subsection{Comparing with existing calibration methods}
To show the efficiency and accuracy of our method, we compared our method with two state-of-the-art autocalibration method \citep{acc2,servomotor} and the gold standard turntable method \citep{truntable}.

\textbf{Conventional turntable method.} This method uses a high-precision turntable to provide a calibration reference. The key parameters of the turntable are 0.0001\degree position accuracy and 0.0001\degree/s angular velocity accuracy. The fixture manufacture precision is $0.02mm$. We use six angular rate method in this study.

\textbf{The accelerometer-aided method.} This method uses the feature that most IMUs include accelerometers, and uses accelerometers to provide a calibration reference for the calibration of the gyroscope. Following the experiment process described in \citep{acc2}, we reproduced the experiment using our gyroscopes. In our experiment, the inclination angle between the device and the desktop is set to 0\degree, 30\degree, and 45 \degree. Different from the original experiment, we did not repeat the experiment multiple times and take average to improve the calibration accuracy, because the experiment time is already as long as ten minutes.

\textbf{The servomotor-aided method.} This method uses a servomotor to provide a reference for calibration and provides a fast and low-cost calibration method. Due to the use of external calibration equipment, this method is not a pure autocalibration method. We have calibrated our gyroscope with reference to the calibration method provided in \citep{servomotor}.

\textbf{Our proposed method} is summarized in Section \ref{sec:process}. 

To compare the efficiency of these method, we also compared the calibration time. We use a stopwatch to calculate the experiment time. Before the start of each experiment, we prepared all the test equipment. After pressing the timing button, we start the experiment, and we stop timing until all the data is collected. We did not include the data processing time, because it usually only takes a few seconds with a well written program.

\subsection{{Error analysis}}

We identified three primary sources of calibration errors, including the biased coursed by the autocalibration model, the discretization error, and physical variations. We discussed them separately as follows:

\textbf{The biased autocalibration model.} The commonly used autocalibration model does not consider the gyroscope measurement error and assume the noise as an Gaussian noise. In practice, the measurements ($m_{x,i,j},m_{y,i,j},m_{z,i,j}$) also contains noise, which lead to the overall noise becomes a combination of Gaussian noise and Chi-square noise,

However, the estimator used, such as Leven-berg-Marquardt algorithm and Nelder-Mead method, cannot handle the noises other than Gaussian noise. This lead to a biased results when estimating the calibration parameters. This problem will gradually disappear as the signal-to-noise ratio increases, such as increase the rotation speed or minimize the gyroscope measurement noise. 
Though we are unable to discuss the speed issue in this article due to the usage of manual rotation, fast rotation speed is consistent with our intention of efficient calibration. 

Regarding the influence of measurement noise, the simulation results in Figure. \ref{errorboxplot} and \ref{errorgyroboxplot} indicates that the calibration error is larger under the condition of high measurement noise. Similar results can be seen from the experiment results (Table \ref{resultsL9} and \ref{resultsICM}). The measurement error of ICM20948 is small, and it has a small calibration error. 

\textbf{The discretization error.} As the proposed method use angle rather than angular velocity as calibration reference, the discretization error is introduced when calculating the angle by angular velocity. This error will decrease as the sampling rate increases. Therefore, we use the highest sampling rate during the self-calibration phase of the device to reduce this error. 

\textbf{Physical variations.} The movement of the two planes in actual operation introduces additional error to calibration references. Theoretically, this error can be eliminated by conduct the calibration process multiple times, but this is contrary to our intention of efficient calibration. Therefore, a relatively solidly desktop should be used for calibration. Compared to the method using a constant speed as the calibration reference, our method does not suffer from the error caused by unstable rotation speed.

\subsection{Results and discussion}

The calibration results of four different calibration methods are summarized in Table. \ref{resultsL9} and \ref{resultsICM}. As we do not know the true scale factors and biases of the testing device, we consider the result of gold standard turntable method as ground truth in the following discussion. 

The absolute error of the proposed method is less than $2.5\times10^{-2}$ for LSM9DS1 and less than $1\times10^{-2}$ for ICM20948. Considering the low repeatability and large measurement noise of the low-cost gyroscopes, the calibration result is considerably accurate. The estimation error of ICM20948 is significantly lower than that of LSM9DS1. We infer that this is because the noise spectral density of ICM20948 is lower. This is also in line with the simulation result, that is, the calibration result of the gyroscope with lower measurement noise has a smaller estimation error. Besides, 
the repeatability of ICM20948 is higher, which means that its parameters change less during the experiments.

\begin{figure}
	\centering
		\includegraphics[height=0.35\textwidth]{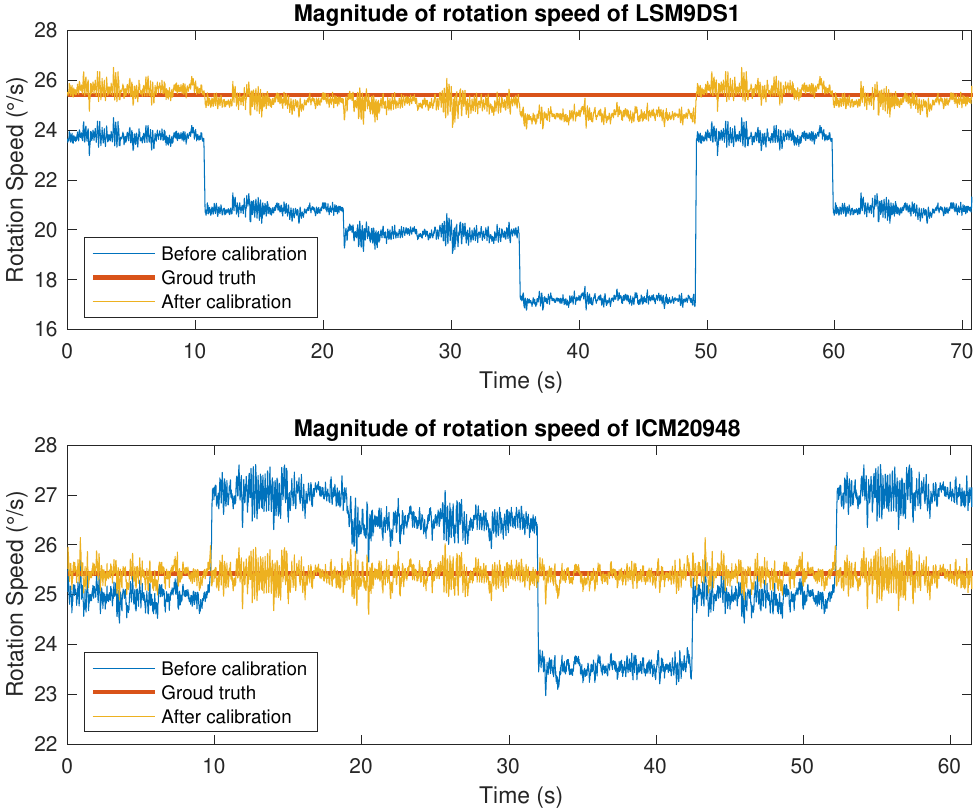}
	\caption{Magnitude of rotation speed before and after calibration. The acceleration and deceleration phases were omitted.}
	\label{mag}
\end{figure}

Compared with the accelerometer-aided method, our method achieves a similar calibration accuracy within one twentieth of the time. Besides, because the accelerometer is used to provide the gyroscope calibration reference, the calibration error of the accelerometer is also added to the gyroscope calibration process. The servomotor-aided method has a fast calibration speed and high calibration accuracy because it has a principle similar to that of a turntable. However, due to the use of external auxiliary equipment, it is not suitable for use in an in-field calibration situation.

To further verify the calibration results, we rotate the above calibrated gyroscope 360 degrees clockwise and counterclockwise along the x, y, z axis at a constant speed of $25.42$\degree/s using a servomotor, and record the raw readings along with the calibration parameters. When plotting, we did not plot the acceleration and deceleration phases. We calculated the magnitude of rotation speed before and after calibration. The magnitude of the rotation speed was calculated by the square root of the sum of squares of their vector components. The results are shown in Figure. \ref{mag}.
The results show that the magnitude of rotation speed after calibration is obviously closer to the ground truth, where the results of ICM20948 almost coincides with the ground truth.

\begin{figure}
	\centering
		\includegraphics[height=0.35\textwidth]{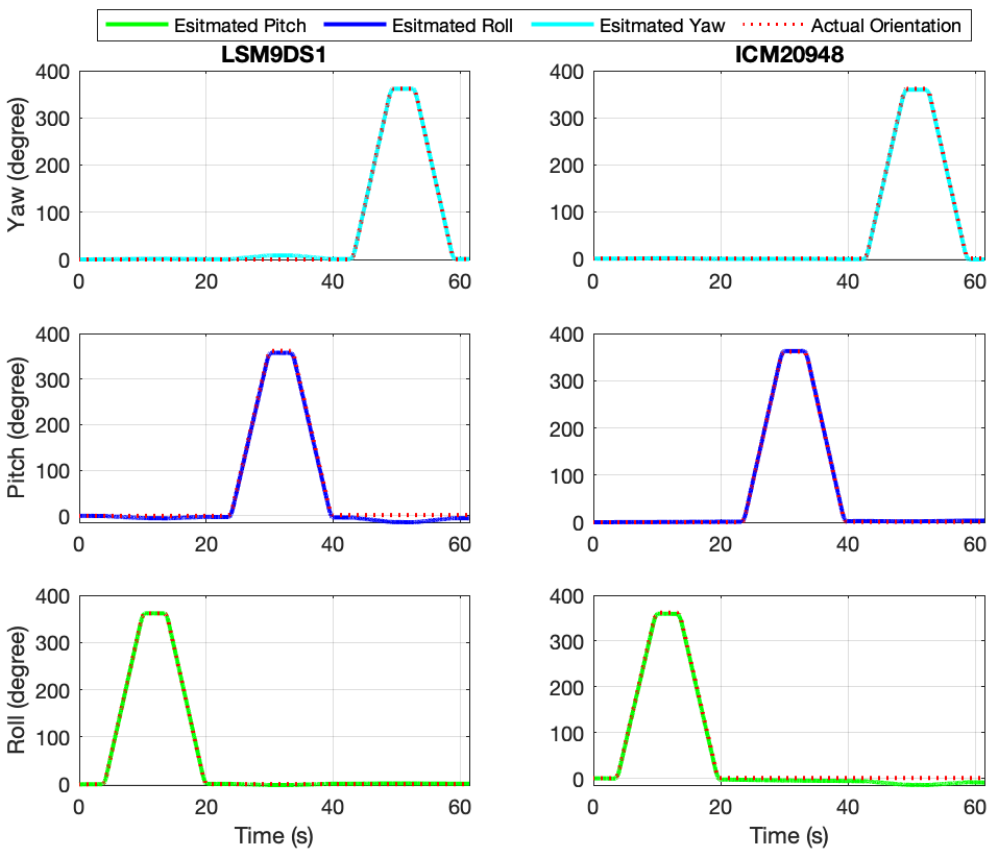}
	\caption{Estimated orientation using calibrated gyroscope readings after calibration from two models of gyroscope. Left: LSM9DS1. Right: ICM20948.}
	\label{oriafter}
\end{figure}

As we discussed in Section \ref{sec:preliminary}, in the transdisciplinary project for monitoring the movements of women with complex pregnancies during labour and childbirth, without calibration, the angle estimation error was significantly high as shown in Figure.\ref{oriBefore}. After using the proposed in-field calibration method, the estimation accuracy has been significantly enhanced. In Figure.\ref{oriafter}, it is clearly showed the ideal trajectories, and the estimated trajectories almost coincide when using the calibrated gyroscope readings.

Overall, the proposed method achieves a relatively high calibration accuracy. It should be emphasized that the entire calibration process only took less than 30 seconds, and any external device is not used during the process.

\section{Conclusion}\label{sec:conclusion}
This paper proposed an efficient in-field gyroscope calibration method for the purpose of developing a wearable monitor to track the movement of a labouring and birthing woman in a hospital birth room. We introduced an efficient in-field calibration method, which is able to readily calibrate the triaxial gyroscopes without additional equipment. 
The feasibility and effectiveness of the proposed approach have been demonstrated via both numerical simulation and experiments.

\bibliographystyle{cas-model2-names}

\bibliography{cas-refs}

\end{document}